\documentclass[12pt]{article}

\usepackage{times}
\usepackage{amssymb}
\usepackage{amsmath}
\usepackage{graphicx}
\usepackage{color}
\usepackage{notes2bib} 
\usepackage{pdfpages}

\makeatletter
\@ifundefined{textcolor}{}
{%
 \definecolor{BLACK}{gray}{0}
 \definecolor{WHITE}{gray}{1}
 \definecolor{RED}{rgb}{1,0,0}
 \definecolor{GREEN}{rgb}{0,1,0}
 \definecolor{BLUE}{rgb}{0,0,1}
 \definecolor{CYAN}{cmyk}{1,0,0,0}
 \definecolor{MAGENTA}{cmyk}{0,1,0,0}
 \definecolor{YELLOW}{cmyk}{0,0,1,0}
 }

\newenvironment{sciabstract}{%
\begin{quote} \bf}
{\end{quote}}

\title{Realization of an atomically thin mirror using monolayer MoSe$_2$}

\author
{Patrick Back,$^{1}$ Aroosa Ijaz,$^{1}$ Sina Zeytinoglu,$^{1}$ Martin Kroner,$^{1\ast}$ Atac Imamo\u{g}lu$^{1\ast}$ \\
\\
\normalsize{$^{1}$Institute for Quantum Electronics, ETH Z\"urich,}\\
\normalsize{CH-8093 Zurich, Switzerland}\\
\\
\normalsize{$^\ast$E-mail:  mkroner@phys.ethz.ch or
imamoglu@phys.ethz.ch} }

\date{}

\begin{document}

\baselineskip24pt

\maketitle

\begin{sciabstract}
Advent of new materials such as van der Waals heterostructures,
propels new research directions in  condensed matter physics and
enables development of novel devices with unique functionalities.
Here, we show experimentally that a monolayer of MoSe$_2$ embedded
in a charge controlled heterostructure can be used to realize an
electrically tunable atomically-thin mirror, that effects $90 \%$
extinction of an incident field that is resonant with its exciton
transition. The corresponding maximum reflection coefficient of $45
\%$  is only limited by the ratio of the radiative decay rate to the
linewidth of  exciton transition and is independent of  incident
light intensity up to $400$ Watts/cm$^2$. We demonstrate that the
reflectivity of the mirror can be drastically modified by applying a
gate voltage that modifies the monolayer charge density. Our
findings could find applications ranging from fast programmable
spatial light modulators to suspended ultra-light mirrors for
optomechanical devices.
\end{sciabstract}

A plethora of ground-breaking experiments have  established
monolayers of transition metal dichalcogenides (TMD) such as
MoSe$_2$ or WSe$_2$ as a new class of two dimensional (2D) direct
band-gap semiconductors~\cite{
Radisavljevic2011,Splendiani2010,Baugher2014,Britnell2013,Xu2014}.
In the absence of free carriers, the lowest energy elementary
optical excitations in TMDs are excitons with an ultra-large binding
energy of $\sim 0.5$~eV~\cite{Chernikov2014}. While encapsulation of
TMD monolayers with hexagonal boron nitride (hBN) leads to a
reduction of the exciton binding
energy~\cite{Thygesen2015,Crooker2016}, it dramatically improves the
optical quality, leading to an exciton line-broadening of $\sim
2$~meV in photoluminescence (PL) or in absorption
measurements~\cite{narrowest-linewidth1,narrowest-linewidth2}.
Remarkably, these narrow linewidths have a dominant contribution
from radiative decay rate; the latter has been determined from
exciton-polariton splitting in open cavity structures to be $\sim
1.5$~meV for MoSe$_2$~\cite{Dufferwiel2015,Sidler2016}. Motivated by
these developments, we previously analyzed the optical response of a
monolayer TMD theoretically and showed that it realizes an
atomically-thin mirror~\cite{Zeytinoglu2017}.

\begin{figure}[!htb]
\centering
\includegraphics{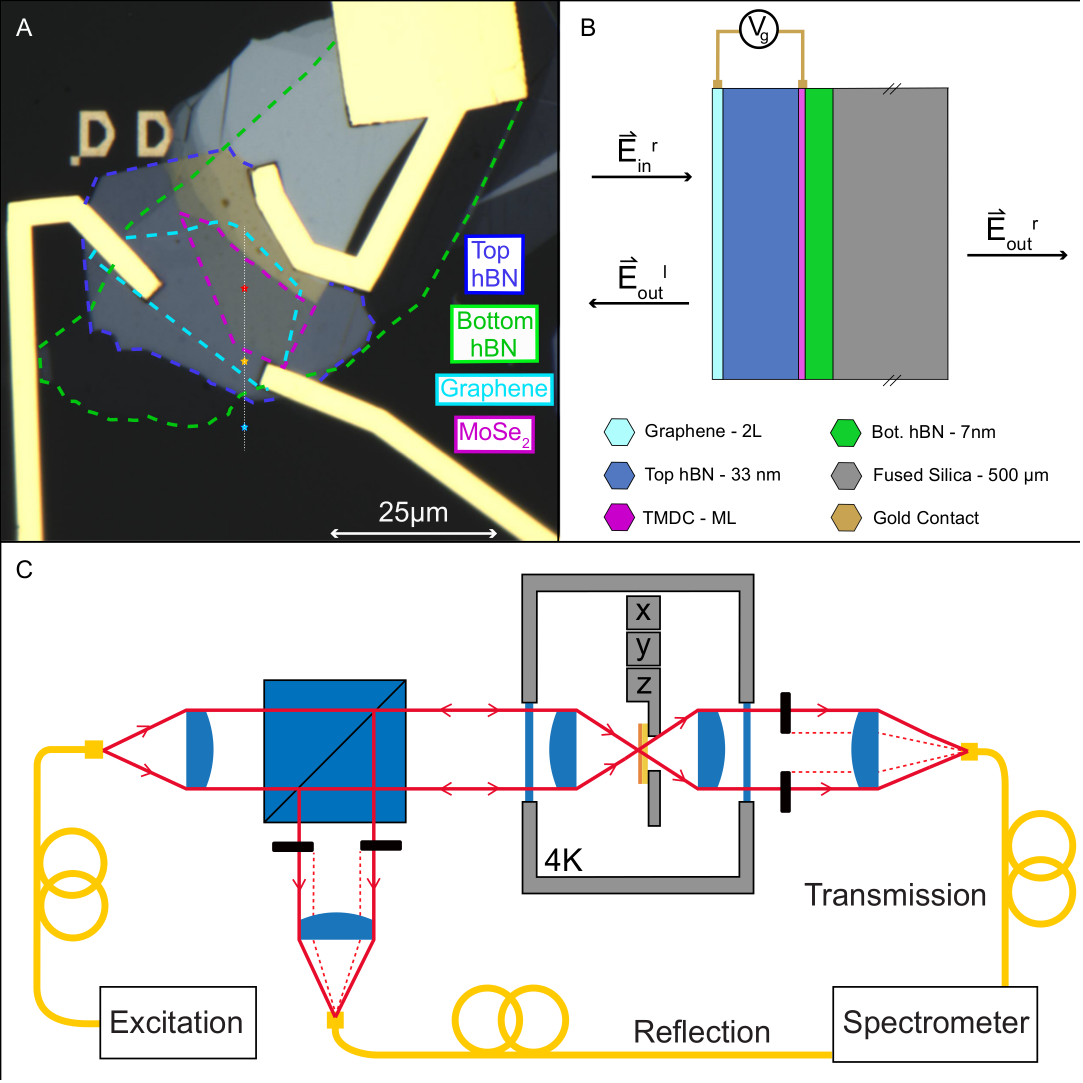}
\caption{{\bf A gate controlled van der Waals heterostructure in a
confocal setup.} (A) Micrograph of the measured heterostructure: the
MoSe$_2$ monolayer is encapsulated in between $33$~nm (top) and
$7$~nm (bottom) thick hBN layers, which are indicated by blue and
green dashed lines. To allow for optical transmission measurements
the heterostructure is transferred on to a transparent 500$\mu$m
thick fused-silica substrate. To achieve charge control of the
MoSe$_2$, the heterostructure is capped by a graphene bilayer. The
graphene and the MoSe$_2$ are electrically contacted by
titanium/gold electrodes. The white line and colored stars indicate
the position of optical measurements on and off the MoSe$_2$ and the
hBN layers. (B) Illustration of the interaction of an incident field
with a MoSe$_2$ monolayer. Optical fields can be characterized as
consisting of right propagating input $E^\mathrm{in}_\mathrm{r}$ and
right and left propagating output modes $E^\mathrm{out}_\mathrm{r}$,
$E^\mathrm{out}_\mathrm{l}$ respectively. (C) The schematic of the
experimental setup. The sample is mounted in a Helium flow cryostat
in between two aspheric lenses in confocal configuration. The sample
can be moved in situ by piezo stepper motors. A collimated
excitation beam is focussed onto the sample by the first lens. The
reflected light is collimated again by the same lens (NA$=0.68$).
The transmitted light is collimated by the second lens (NA=0.55). By
reducing the diameter of the collection beam we can reduce the
effective NA of the detection optics, thereby reducing the spread in
the in-plane momentum k of the detected photons. Since the
excitation paths remain unchanged, we only excite a small area of
the sample which guarantees little excess inhomogeneous broadening
due to disorder of the flake even for a small detection NA.}
\label{fig:1}
\end{figure}

Here, we present experiments demonstrating that an hBN encapsulated
MoSe$_2$ monolayer has a reflection coefficient exceeding $40 \%$
and an extinction of resonant transmitted light of $90 \%$. The
combined reflection and transmission measurements allow us to
conclude that the fraction of scattered light, stemming from
inhomogeneous broadening of the exciton resonance, is less than
$50\%$. Figure~\ref{fig:1}A shows a micrograph of the charge tunable
heterostructure we have studied: a $9 \mu$m by $4 \mu$m MoSe$_2$
monolayer is sandwiched between two hBN layers. The heterostructure
is placed on top of a transparent fused-silica substrate and capped
with bilayer graphene. A gate voltage $V_\mathrm{g}$ applied between
bilayer graphene and the MoSe$_2$ layer allows for tuning the
electron density $n_\mathrm{e}$~\cite{Ross2013,Chernikov2015} and
thereby modifying the resonance frequency as well as the nature of
the elementary optical excitations~\cite{Sidler2016}.

Figure~\ref{fig:1}B depicts the outline of our experiments. We use
near-resonant light to probe the excitonic excitations. To the
extent that the MoSe$_2$ monolayer and its environment is
homogeneous, in-plane momentum $k$ is a good quantum number for both
exciton and radiation field modes. In this limit, we use the
input-output formalism of quantum optics to express the outgoing
right-propagating electric field mode
E$_\mathrm{out}^\mathrm{r}(\omega)$ (Fig.~\ref{fig:1}B) in terms of
the incident field modes E$_\mathrm{in}^\mathrm{r}(\omega)$ and
E$_\mathrm{in}^\mathrm{l}(\omega)$~\cite{Zeytinoglu2017}
\begin{align}
E_\mathrm{out}^\mathrm{r}(\omega) &=  \frac{\Gamma/2}{\Gamma/2 -
i(\omega - \omega_\mathrm{exc})} [ (1- \eta)
E_\mathrm{in}^\mathrm{r}(\omega) - \eta
E_\mathrm{in}^\mathrm{l}(\omega)], \label{eq:IandO}
\end{align}
where $\omega_\mathrm{exc}$ and $\Gamma$ denote the transition
frequency and the radiative decay rate of the excitons.  $\eta =
\Gamma/\gamma_\mathrm{tot}$ gives the ratio of radiative decay rate
to the total (Lorentzian) exciton line broadening
$\gamma_\mathrm{tot}$. In our experiments $
E_\mathrm{in}^\mathrm{r}(\omega) = E_\mathrm{inc} \neq 0$ whereas
$E_\mathrm{in}^\mathrm{l}(\omega)= 0$;  in this limit the reflected
and transmitted fields are given by $E_\mathrm{refl} =
E_\mathrm{out}^\mathrm{l}(\omega)$ and $E_\mathrm{trans} =
 E_\mathrm{out}^\mathrm{r}(\omega)$, respectively. The minimum and maximum
(resonant) intensity transmission and reflection coefficients are
given by $T_\mathrm{min} = (\gamma_\mathrm{tot} -
\Gamma)^2/\gamma_\mathrm{tot}^2$ and $R_\mathrm{max} =
\Gamma^2/\gamma_\mathrm{tot}^2$. Our formulation does not take into
account small asymmetry in radiative decay rates into left and right
propagating field modes stemming from the fused-silica substrate.

We measure the transmission and reflection spectrum of the
heterostructure in a cryogenic transmission microscope with free
space optical access as shown in Fig.~\ref{fig:1}C. The incident
light is focussed to a diffraction limited spot with diameter
$\approx 700\mathrm{nm}$. For the excitation we use either a broad
band light emitting diode (LED) centered at 780nm or a tunable
single-mode Ti:S laser. Due to losses on the windows of the cryostat
and the finite coupling/detection efficiency we can not directly
measure the absolute reflected and transmitted power. Instead, we
use a reference reflection and transmission spectrum of the incident
light by moving our excitation spot off the MoSe$_2$ layer or off
the heterostructure to the fused-silica layer.

\begin{figure}[!htb]
\centering
\includegraphics{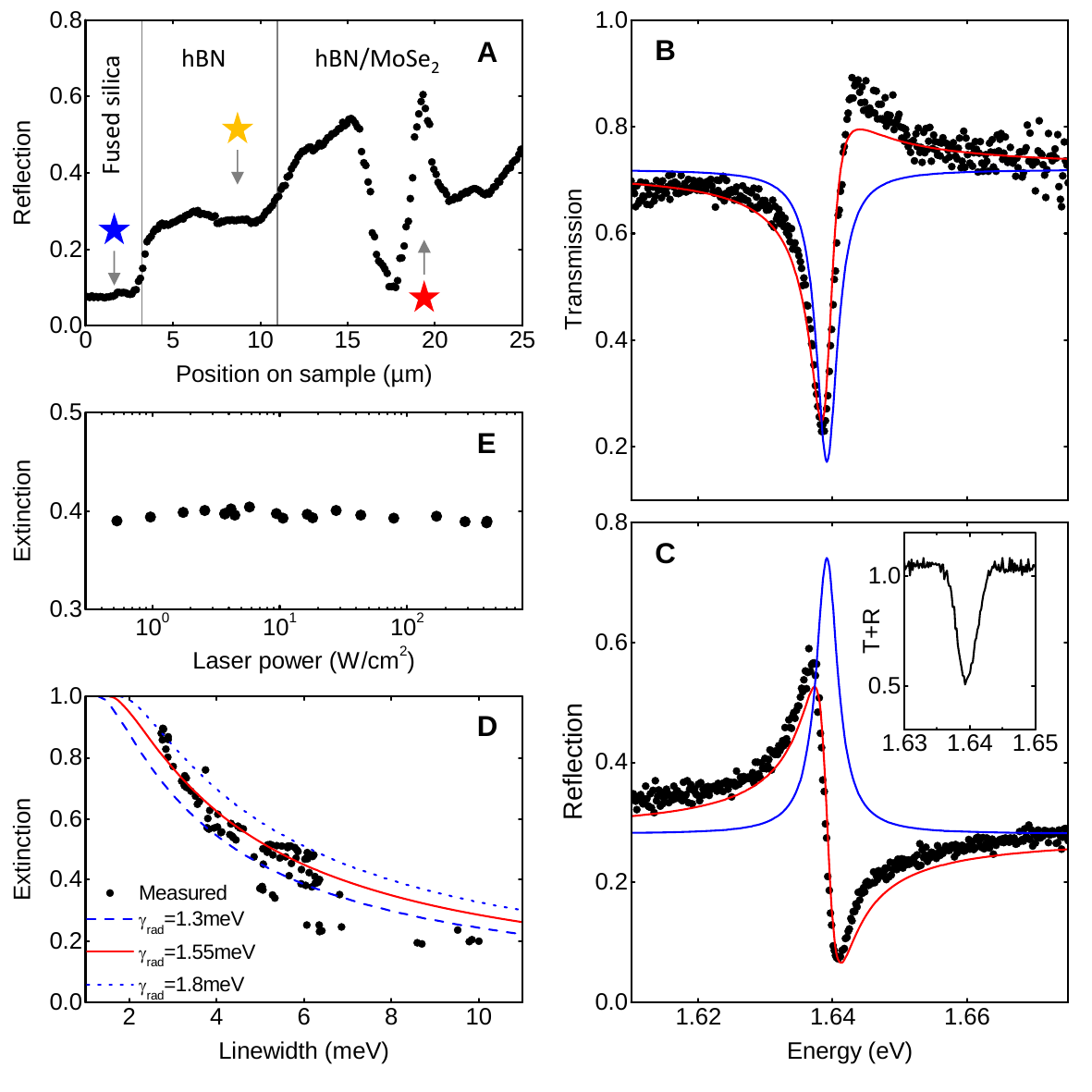}
\caption{{\bf Resonant optical response of a MoSe$_2$ monolayer.}
(A) The measured reflection coefficient at a photon energy of
$1.64$eV across the white line indicated in Fig.\ref{fig:1}A. The
reflection increases from $8 \%$ in the fused-silica region (blue
star) to about $28 \%$ in the region having both top and bottom hBN
layers as well as bilayer graphene (yellow star). The strong
variations in reflection between 12$\mu$m and 25$\mu$m stems from
the position dependence of the exciton resonance of the MoSe$_2$.
The incident photons are resonant with the exciton transition at the
position indicated by the red star. (B) Transmission as function of
photon energy of the MoSe$_2$/hBN heterostructure measured at the
position indicated by the red star with a detection arm numerical
aperture of $0.1$. The black dots represent the normalized data (see
Supplementary Materials). The red line is a fit to the data using a
dispersive Lorentzian model yielding  $\gamma_\mathrm{tot}=3.6$meV.
The blue line shows the corresponding Lorentzian, indicating an
extinction coefficient of $0.76$. (C) Reflection as function of
photon energy measured under identical conditions as in (B).  The
fit allows us to extract resonant reflection enhancement by a factor
of $1.67$, indicating that the maximum reflectivity of the MoSe$_2$
monolayer is $46 \%$. The inset shows the sum of the transmission
and reflection spectra. (D) Demonstration of linewidth dependence of
maximal extinction. The black dots show the extracted extinction on
resonance ($1-T_\mathrm{min}$) as function of the measured linewidth
$\gamma_\mathrm{tot}$. The lines indicate the dependence of the
extinction as function of $\gamma_\mathrm{tot}$ for different
radiative decay rates $\Gamma$. (E) Extinction of a single-mode
laser resonant with the exciton transition as function of its
intensity.} \label{fig:2}
\end{figure}

Figure~\ref{fig:2}A shows the intensity reflection coefficient of
light with a photon energy of $1.64$~eV as we move the
excitation/detection spot along the white line indicated in
Fig.~\ref{fig:1}A. In the region around the blue spot, we measure
the reflectivity of fused-silica layer; the extracted reflection
coefficient of $8 \pm 1\%$ (see Supplementary Materials) is in good
agreement with $9.57 \%$ reflection we would expect from a thick
fused-silica layer. Moving to the orange spot where we have the
graphene/hBN heterostructure without the MoSe$_2$ monolayer, we
observe a sizeable increase in reflection coefficient to about $28
\pm 5 \%$. Finally, reflection of the full heterostructure around
the red spot exhibits very large variations in reflection
coefficient ranging from $10 \%$ to more than $60 \%$, demonstrating
the ultra-strong optical response of monolayer MoSe$_2$. Spatial
variations in exciton resonance frequency, most likely stemming from
inhomogeneous strain profile, result in the observed position
dependent reflection coefficient of incident photons at a fixed
frequency.

Figure~\ref{fig:2}B shows the normalized transmission spectrum
obtained at the red spot, using a detection path numerical aperture
(NA) of $\sim 0.1$. Even though the strong extinction of resonant
light is evident, the dispersive nature of the transmission spectrum
stemming from optical interference effects, does not allow us to
read $T_\mathrm{min}$ from the data directly. To extract the maximal
extinction, we fit the data with a dispersive Lorentzian model (red
curve). We find that a good fit is obtained for
$\gamma_\mathrm{tot}= 3.6$~meV yielding an extinction coefficient of
$1 - T_\mathrm{min} = 0.76\pm 0.15$. We emphasize that extinction
factors of this magnitude were previously only reported for transmon
qubits coupled to superconducting microwave
waveguides~\cite{Wallraff2013}.


The reflection spectrum corresponding to the transmission data
obtained at the red spot is depicted in Fig.~\ref{fig:2}C. The
stronger asymmetry in reflection originates from an optical
interference between two contributions with comparable magnitude --
the excitonic emission and the incident light reflected from the
heterostructure. The parameters used in the dispersive Lorentzian
fit (red curve) to the reflection data using the same resonance
energy and $\gamma_\mathrm{tot}$ as for the transmission spectrum,
allows us to determine the maximal reflection coefficient of the
MoSe$_2$ layer to be $(46 \pm 9) \%$. Remarkably, the sum of
reflection and transmission, normalized to the corresponding value
in the region around the orange spot, yields a symmetric but
non-Lorentzian dip at the exciton resonance energy
(Fig.~\ref{fig:2}C inset): the reduction of the total light
intensity on resonance is a consequence of non-radiative line
broadening and confirms that losses due to scattering into high $k$
field modes or due to non-radiative recombination is less than $50
\%$.

The aforementioned spatial inhomogeneity of the exciton resonance
results in a variation of not only the exciton resonance energy but
also its linewidth across the sample. Figure~\ref{fig:2}D shows the
dependence of the extracted resonant extinction coefficient on the
linewidth of the exciton resonance, obtained by measuring
transmission at different spots. We find that the broadened
lineshapes are reasonably well fit by dispersive Lorentzians
allowing us to compare the dependence of maximal extinction on total
exciton linewidth ($\gamma_\mathrm{tot}$ to the theoretical value $1
- (\gamma_\mathrm{tot} - \Gamma)^2/\gamma_\mathrm{tot}^2$. This
comparison in turn allows us to determine the radiative decay rate
to be in the range $1.4$~meV $< \Gamma <$ $1.8$~meV, which is in
good agreement with the values previously determined from
normal-mode splitting of
exciton-polaritons~\cite{Dufferwiel2015,Sidler2016}. We remark that
at the locations yielding the highest observed extinction factor of
$\sim 0.9$, we extract consistently higher values of $\Gamma$.
However, at these points even the transmission spectrum exhibits
strong asymmetry, rendering the extracted values less reliable.

Since the exciton-exciton interactions are proportional to the Bohr
radius $a_\mathrm{B}$~\cite{Ciuti1998}, we would expect strongly
bound TMD excitons with $a_\mathrm{B} \simeq 1$~nm to have a linear
response to the incident field, provided that the induced exciton
density $n_\mathrm{exc} \ll 1 \times 10^{12}$~cm$^{-2}$. This
argument in turn suggests that the response of this atomically-thin
mirror should be independent of the incident light intensity
$I_\mathrm{L}$ provided that $I_\mathrm{L} \ll 1 \times
10^{5}$~Wcm$^{-2}$. To verify this prediction, we used a single-mode
Ti:S laser tuned into resonance with the MoSe$_2$ exciton transition
and monitored the dependence of extinction on the laser intensity.
Figure~\ref{fig:2}E shows that the extinction indeed remains
unchanged as we vary the incident intensity by 4 orders of magnitude
from $0.5$ W/cm$^2$ to $400$ W/cm$^2$. Verification of the
theoretically predicted unusual saturation characteristics of TMD
mirrors is likely to require pulsed laser
excitation~\cite{Zeytinoglu2017}.


Due to strong exciton-electron interactions, the nature of
elementary optical excitations can be drastically modified by
applying $V_\mathrm{g}$ that modifies the free electron (or hole)
density ($n_\mathrm{e}$) in the TMD monolayer~\cite{Sidler2016}. A
typical $V_\mathrm{g}$ dependent transmission spectrum is shown in
Fig.~S4 (Supplementary Materials): as $n_\mathrm{e}$ is increased a
new attractive exciton-polaron resonance that is red detuned by
$25$~meV with respect to the bare exciton transition emerges. The
abrupt blue-shift and broadening of the exciton resonance for
$V_\mathrm{g}
> -2$~V on the other hand can be understood as the transition from a
bare exciton into a repulsive-exciton-polaron resonance. In the
context of our work, the $V_\mathrm{g}$ dependence of the optical
response of the MoSe$_2$ monolayer indicates that the
atomically-thin mirror can be electrically tuned.

\begin{figure}[!htb]
\centering
\includegraphics{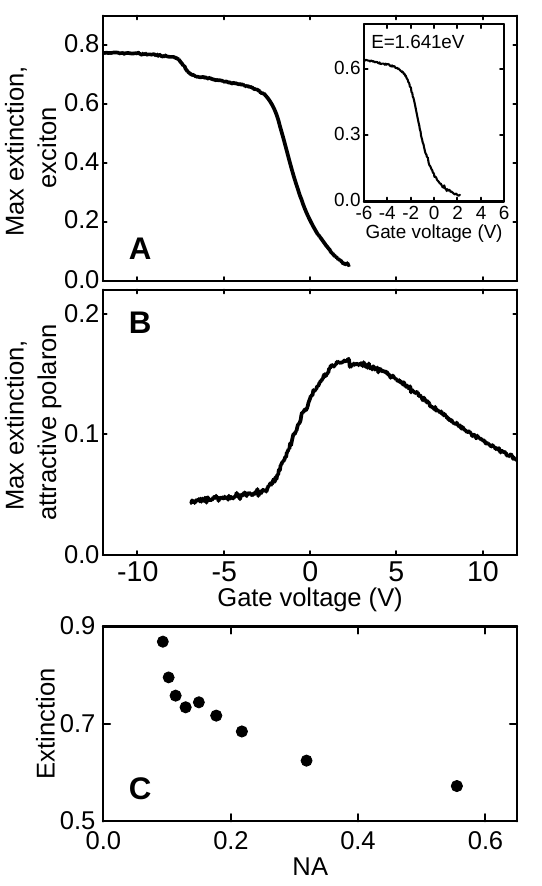}
\caption{{\bf Electrical control of an atomically-thin mirror.} (A)
Gate voltage ($V_\mathrm{g}$) dependence of the maximal extinction
of transmitted light using the exciton resonance. Injection of free
carriers reduces the exciton oscillator strength and increases
linewidth, leading to a sharp $V_\mathrm{g}$ dependent drop in
transmission. The inset shows the extinction of incident photons
with a fixed energy of $1.641$eV as function of $V_\mathrm{g}$
demonstrating that transmission can be changed from $10 \%$ to $60
\%$ by changing $V_\mathrm{g}$ by $\sim 2$~V. (B) $V_\mathrm{g}$
dependence of maximal extinction at the attractive exciton-polaron
resonance: the extinction increases with increasing electron density
until $V_\mathrm{g} \sim 2$~V due to oscillator strength transfer
from the exciton resonance. For $V_\mathrm{g} > 2$~V, excess line
broadening suppresses the resonant enhancement of optical response.
(C) Measured maximal extinction as function of the NA of the
detection optics. Increasing NA leads to an increase of the
linewidth and a corresponding reduction of extinction. The
observations are consistent with electron-hole exchange interaction
induced splitting of s- and p-polarized excitonic resonances which,
in the absence of polarization selection, leads to line broadening.}
\label{fig:3}
\end{figure}

Figure~\ref{fig:3}A shows the $V_\mathrm{g}$ dependence of  maximal
extinction coefficient obtained at exciton or repulsive-polaron
resonance. The increase of minimum transmission from $\sim 20 \%$
down to $90 \%$ is due to a combination of electron-exciton
interaction induced line broadening and oscillator strength transfer
to attractive-polaron. If we instead focus on the response to
incident photons at a specific energy, we find an even sharper drop
in extinction from $0.6$ down to $0.1$ upon increasing
$V_\mathrm{g}$ from $-2$~V to $0$~V, demonstrating electrical
control of an atomically-thin mirror (Fig.~\ref{fig:3}A inset). With
low resistance Ohmic contacts to the MoSe$_2$ monolayer, it should
be possible to exploit the strong $V_\mathrm{g}$ dependence of
excitonic response to realize fast switching of mirror
transmission/reflection potentially on sub-ns
timescales~\cite{Krasnozhon}.

Due to the aforementioned oscillator strength transfer, maximal
extinction at the attractive-polaron resonance first increases to
about $0.2$ before decreasing due to line broadening
(Fig.~\ref{fig:3}B). Even though the optical response of the
attractive polaron is relatively modest, its strong dependence on
the valley polarization of electrons and its large g-factor
exceeding 15~\cite{Back2017} render it an excellent candidate for
realization of chiral optical devices.

The experiments shown in Fig.~\ref{fig:2}A-D as well as in
Fig.~\ref{fig:3}A-B have been obtained by focusing the incident
light onto a diffraction limited spot using a lens with NA $= 0.68$
and then collecting the transmitted or reflected light with an
NA~$\sim 0.1$. This allowed us to monitor the optical response of
excitons with $k \sim 0$. Figure~\ref{fig:3}C shows the dependence
of maximal extinction for several values of the transmission path
NA, close to the same spot we used to obtain the data depicted in
Fig.~\ref{fig:2}B. We observe that as we increase transmission NA
from $0.1$ to $0.55$, the extinction drops from $0.9$ down to $0.6$
while the exciton linewidth increases from $2.5$~meV to $3.5$~meV.
We attribute this strong NA dependence to electron-hole-exchange
interaction induced modification of the exciton spectrum: it has
been theoretically predicted that due to this interaction, the
p-polarized excitons would have a Dirac-cone-like dispersion,
leading to an energy splitting of s- and p-polarized excitons of
order $3$~meV for $k \sim
\omega_\mathrm{exc}/c$~\cite{Wangyao2014,Louie2015}. To the best of
our knowledge, the NA-dependent increase in linewidth that we report
here provides the first direct evidence for this striking
theoretical prediction.

The strong optical response of a TMD monolayer as demonstrated in
our experiments opens up new avenues for photonics. On the one hand,
the combination of ultra-light mass and high reflectivity suggests
that these atomically-thin mirrors could revolutionize the
performance of optomechanical mass and force sensors. On the other
hand, the possibility to drastically modify reflection on
ultra-short time scales on sub-wavelength length scales using
applied electric fields could open up new perspectives for digital
mirror devices or spatial light modulators. Last but not least, the
valley degree of freedom of excitons and exciton-polarons can be
used to realize chiral mirror devices by introducing a ferromagnetic
monolayer next to the optically active TMD layer~\cite{Huang}.

\section*{Acknowledgments}
This work is supported by an ERC Advanced investigator grant
(POLTDES), Spin-NANO Marie Sklodowska-Curie Grant Agreement No.
676108 and NCCR QSIT.

\newpage

\clearpage


\section*{Supplementary material}

# Supplementary Materials for
# Realization of an atomically thin mirror using monolayer MoSe$_2$

Patrick Back,[1] Aroosa Ijaz,[1] Sina Zeytinoglu,[1] Martin Kroner,[1*] Atac Imamoğlu[1*]

[1]Institute for Quantum Electronics, ETH Zürich,

CH-8093 Zurich, Switzerland

*To whom correspondence should be addressed;

E-mail: mkroner@phys.ethz.ch or imamoglu@phys.ethz.ch

## I. MEASUREMENT SETUP

The sample investigated in the main text is a monolayer MoSe$_2$ flake, encapsulated in between two hBN layers. The monolayer is gate tunable thanks to bilayer graphene placed on the top hBN. The sample was transfered onto a 0.5 mm thick transparent fused-silica substrate using a dry transfer method [1]. Using atomic force microscope (AFM) measurements we estimate the surface roughness of the substrate to be about $\Delta t = 10$ nm. The hBN layer between the MoSe$_2$ flake and the bilayer graphene has a thickness of $t_1 \simeq 33$ nm and the hBN between the MoSe$_2$ and the fused-silica has a thickness of $t_2 \simeq 7$ nm. An AFM image of the main text sample and the corresponding line-cut used to determine the thicknesses is shown in Fig. S1. The sample is contacted using titanium/gold and can be doped by applying a voltage between the bilayer graphene and the MoSe$_2$ monolayer. For the measurements, the device was mounted in a flow cryostat with free space optical front and back access.

## II. CAPACITIVE MODEL FOR THE FERMI ENERGY

The electron density $n_e$ of the gated monolayer MoSe$_2$ sample can be tuned by applying a gate voltage V$_g$ between the bilayer graphene and the MoSe$_2$ monolayer. From Fig. S2, we estimate the gate voltage for which we start populating the lowest energy conduction band to be V$_g = -8$ V. For the gate dependent extinction measurements shown in Fig. S4, the injected $n_e$ appears to remain constant at a value $\ll 1 \times 10^{12}$ cm$^{-2}$ in the gate voltage range $-8$ V $<$ V$_g < -2$ V: the characteristic sharp blue-shift of the exciton resonance, along with the substantial oscillator strength transfer to the attractive polaron resonance is observed in the range V$_g > -2$V in Fig. S4.

The capacitance per unit area C/A between bottom gate and sample is given by:

$$\frac{C}{A} = \left(\frac{t_{\text{hBN}}}{\epsilon_{\text{hBN}}\epsilon_0} + \frac{1}{e^2 \text{D}(E)}\right)^{-1}, \tag{1}$$

with $\text{D}(E)$ denoting the density of states of the monolayer. The structure used for the experiments shown in the main text has an hBN layer with thickness $t_{\text{hBN}} \simeq 33$ nm and static (perpendicular) dielectric constant $\epsilon_{\text{hBN}} \approx 3$. Therefore, for this structure $L/\epsilon = t_{\text{hBN}}/\epsilon_{\text{hBN}} = 11$ nm. The second term in the expression, the quantum capacitance, can be neglected as long as the Fermi energy lies within the conduction or valence band.

### III. PHOTOLUMINESCENCE AND TRANSMISSION MEASUREMENTS

We took gate dependent photoluminescence (PL) measurements from the sample depicted in the main text. A typical gate dependent PL measurement is shown in Fig. S2. The excitation laser wavelength was 714 nm. We note that the resonance energy/wavelength varies across the sample. This can be verified by comparing the resonance wavelengths of Fig. S2 and S3 that were obtained on two different spots on the same sample.

Figure S4 shows a color coded plot of back gate (BG) voltage dependent transmission spectrum. For low BG voltages when the sample has low electron densities, there is high extinction at the exciton resonance frequency. By increasing the BG voltage, $n_e$ is increased and we reach a regime where we get sizeable extinction at the attractive exciton-polaron resonance.

### IV. NORMALIZATION OF REFLECTION AND TRANSMISSION SPECTRA

In order to obtain the reflection and transmission coefficients of the sample we measure both integrated transmitted $I_\text{t}$ and reflected $I_\text{r}$ power across the sample (along the white line shown in Fig. 1A of the main text). From these measurements we extract values for the transmitted (reflected) power on the the fused-silica $I_{\text{t,fs}}$ ($I_{\text{r,fs}}$) and the hBN $I_{\text{t,hBN}}$ ($I_{\text{r,hBN}}$) but off the MoSe$_2$ flake. We assume that

$$1 = R_{\text{fs}} + T_{\text{fs}} = \alpha_{\text{r}} I_{\text{r,fs}} + \alpha_{\text{t}} I_{\text{t,fs}}, \tag{2}$$
$$1 = R_{\text{hBN}} + T_{\text{hBN}} = \alpha_{\text{r}} I_{\text{r,hBN}} + \alpha_{\text{t}} I_{\text{t,hBN}}. \tag{3}$$

Here, $\alpha_\text{r}$ and $\alpha_\text{t}$ account for the losses of the detection system in the reflection and transmission arm, respectively. Our experimental accuracy is insufficient to measure directly the influence of the bilayer graphene on the transmitted or reflected power on the hBN.

We find $R_{\text{fs}} = 0.076 \pm 0.015$ and $R_{\text{hBN}} = 0.28 \pm 0.06$ and the corresponding values for the transmission coefficients. $R_{\text{fs}}$ corresponds well with the expected reflection coefficient of the fused-silica substrate (0.096). A matrix transfer calculation of reflection coefficient of the hBN/fused-silica layer structure using the layer thicknesses obtained by AFM however requires a refractive index of $n \sim 2.5$ for the hBN to yield the experimentally observed value. The measured reflection coefficient for an excitation energy of 1.6391eV is plotted in Fig. 2A of the main text.

In order to obtain the spectrum of the reflection and transmission coefficients of the MoSe$_2$/hBN/fused-silica heterostructure we record a transmission $T_{\text{on}}$ and reflection $R_{\text{on}}$ spectrum, for a given electron density or detection NA on the MoSe$_2$/hBN/fused-silica heterostructure (position indicated by the red dot in Fig. 1A of the main text). Typical spectra obtained this way are plotted using black dots in Fig. S5A for transmission and Fig. S5B for reflection, measured at $V_g = 10V$ and $NA = 0.1$. For normalization we take a reference spectrum off the MoSe$_2$ monolayer but still on the bilayer graphene, the top hBN and the bottom hBN layer (orange star in Fig. 1A of the main text) in transmission $T_{\text{off}}$ and reflection $R_{\text{off}}$ (red dots in Fig. S5 A and B). We calculate the TMD monolayer transmission and reflection coefficient at each photon energy as

$$T = \frac{T_{\text{on}}/T_{\text{off}}}{T_{\text{hBN}}}, \tag{4}$$

$$R = \frac{R_{\text{on}}/R_{\text{off}}}{R_{\text{hBN}}}. \tag{5}$$

## V. RESULTS FROM OTHER SAMPLES

We carried out transmission measurements on 4 different samples. The results across all samples differed only in magnitude depending on exciton linewidth. Fig. S6 shows the 68% extinction seen in one of the samples without charge control. Although the line widths observed in this sample were broader than the main sample, the general relation between the amount of extinction and resonance linewidth (see inset of Fig. S6 ) is similar to that shown in Fig. 2D of the main text.

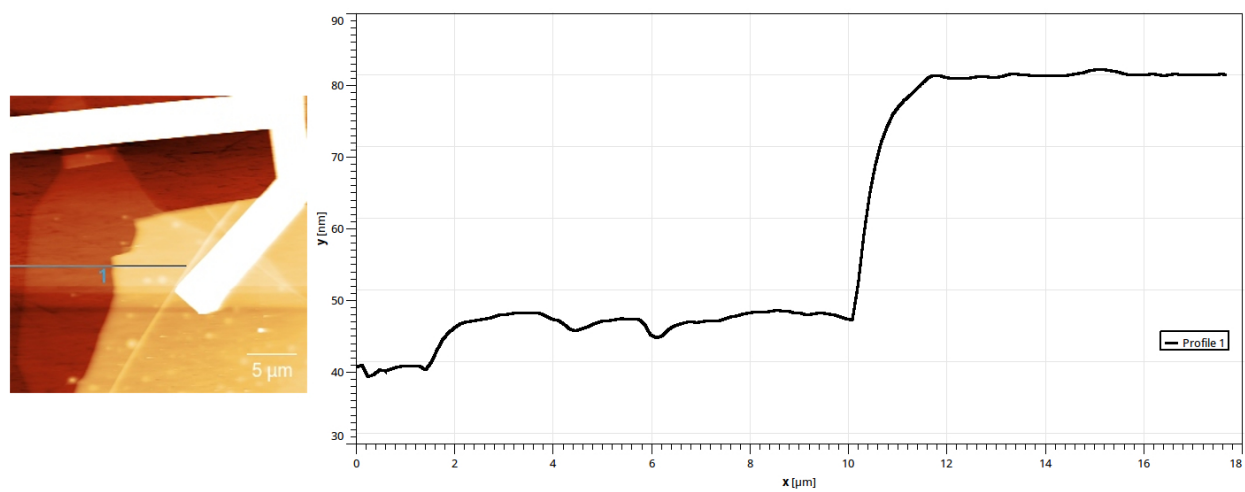

FIG. S1: AFM image taken on the sample of the main text. From the line profile, we estimate the bottom and top hBN layers to have a thickness of $\simeq 7$ nm and $\simeq 33$ nm, respectively.

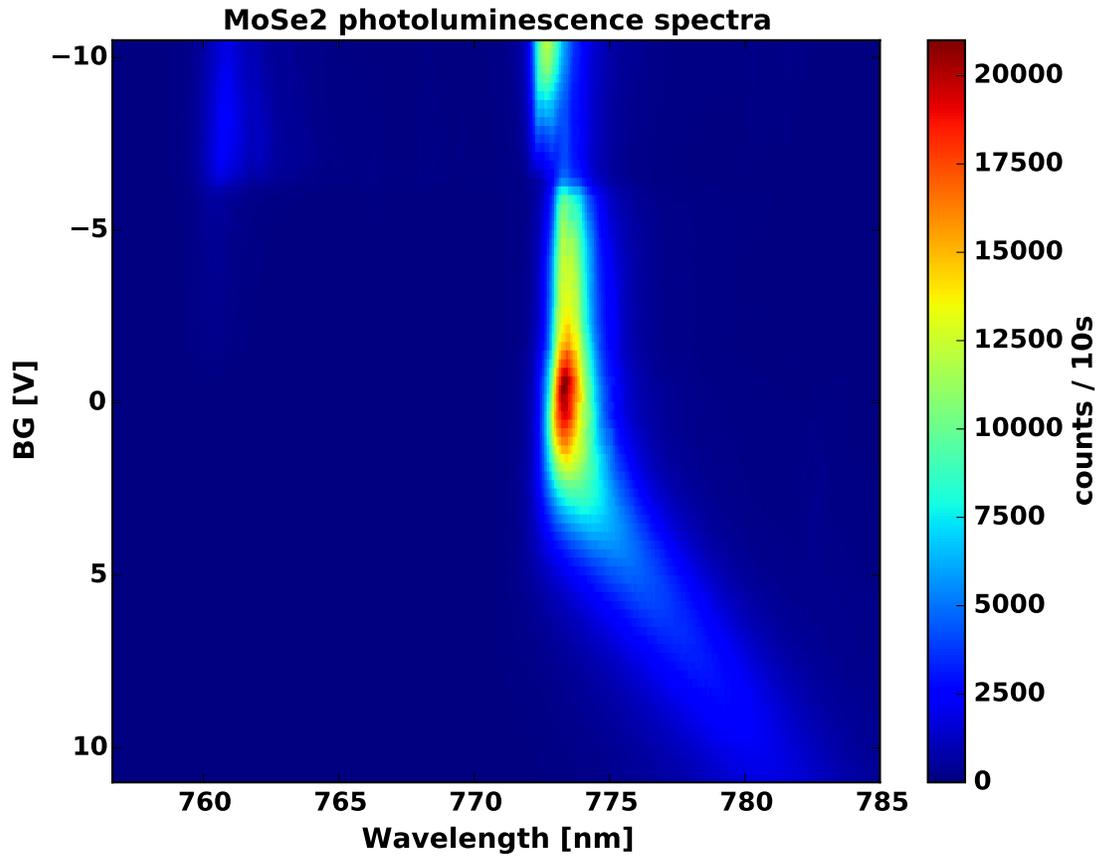

FIG. S2: Electron density dependent photoluminescence measurement of monolayer MoSe$_2$. We observe the PL from exciton for low gate voltages $V_g$ and from attractive polaron for all gate voltages. As $n_e$ is increased (higher $V_g$), the attractive polaron resonance broadens and shifts to higher wavelengths.

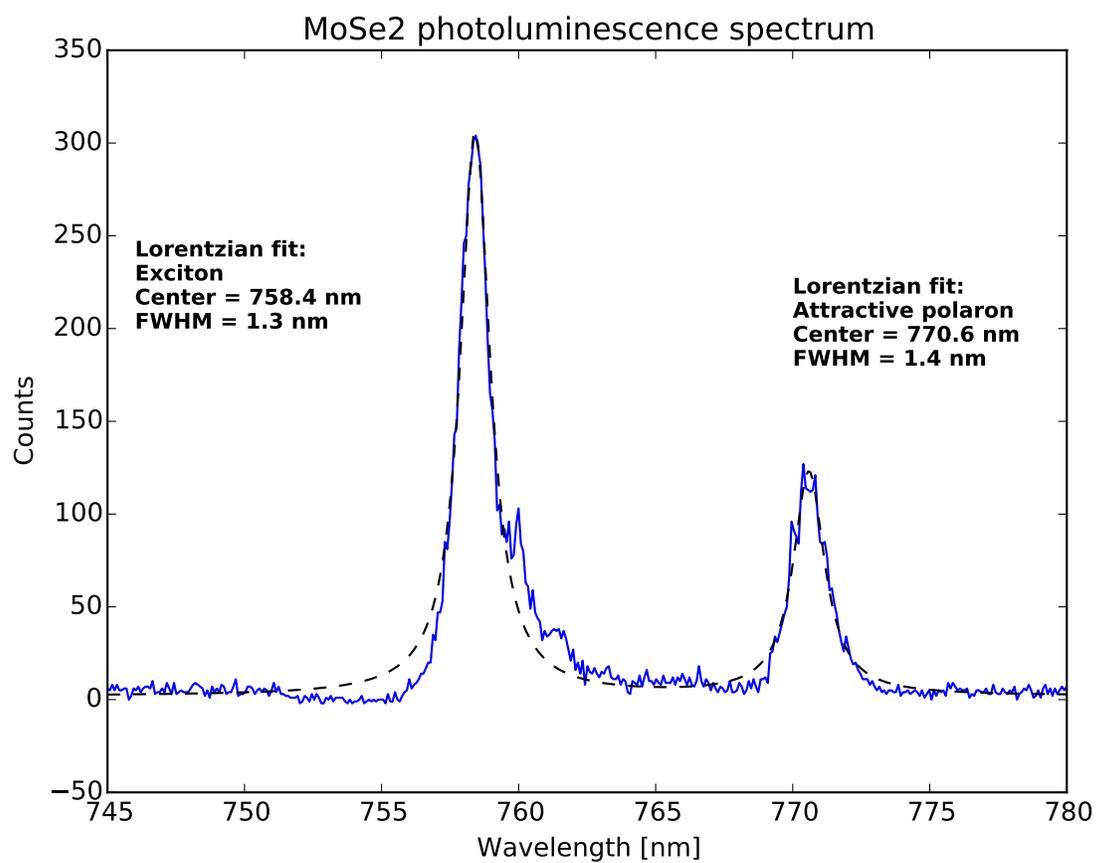

FIG. S3: PL spectrum obtained at a spot yielding the narrowest linewidth on the sample investigated in the main text. The gate voltage was chosen to maximize the exciton PL. By fitting the exciton resonance with a Lorentzian, we extract a line width of $\simeq 2.8$ meV.

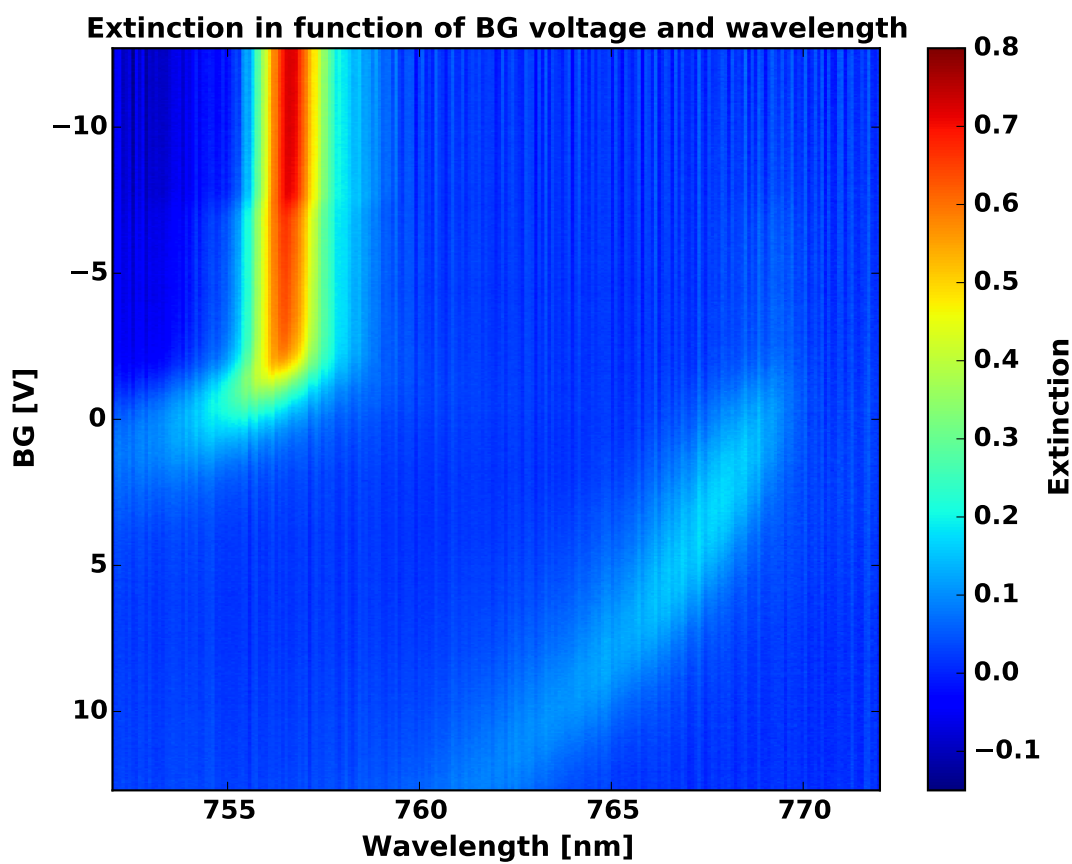

FIG. S4: Transmission spectrum of the heterostructure obtained using a broadband LED source.

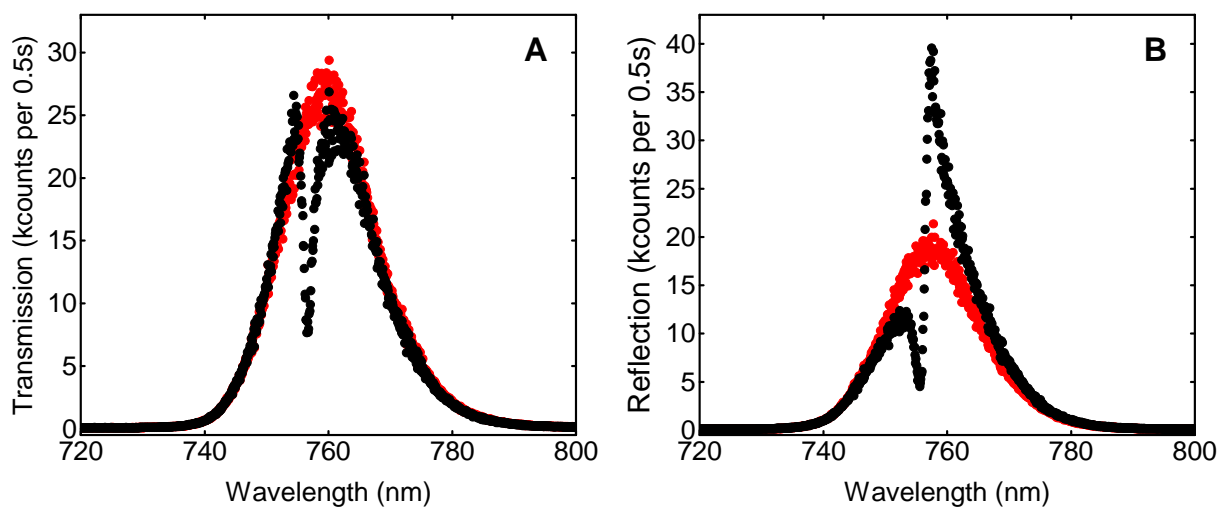

FIG. S5: Measured broadband LED transmission and reflection spectra on (black dots) and off (red dots) the MoSe$_2$ monolayer. The off MoSe$_2$ monolayer spectrum is recorded on a part of the sample having both hBN layers and the bilayer graphene.

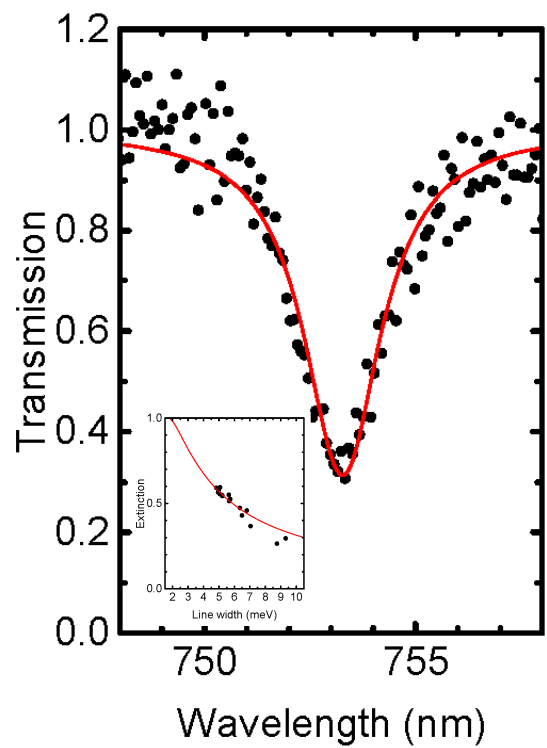

FIG. S6: The extinction of transmitted light measured on a hBN/MoSe$_2$/hBN heterostructure on fused silica sample with full collection NA = 0.68 (without charge control). The fit corresponds to 68% extinction and a line width of 4.8 meV. The inset shows the relationship between the amount of extinction and line width measured on different spots of this sample.



\end{document}